\documentclass{ws-procs9x6}
\usepackage{graphicx}
\def\@fmsl@sh#1#2#3{\m@th\ooalign{$\hfil#1\mkern#2/\hfil$\crcr$#1#3$}}
 \def\eq#1\en{\begin{equation}#1\end{equation}}
\def\s[#1,#2]{[#1\stackrel{\star}{,}#2]}
\def\sx[#1,#2]{[#1\stackrel{\star_{x}}{,}#2]} \def\pp#1{\partial_#1}

\input{epsf} 

\begin{document}

\makeatletter
\def\fmslash{\@ifnextchar[{\fmsl@sh}{\fmsl@sh[0mu]}}
\def\fmsl@sh[#1]#2{%
  \mathchoice
    {\@fmsl@sh\displaystyle{#1}{#2}}%
    {\@fmsl@sh\textstyle{#1}{#2}}%
    {\@fmsl@sh\scriptstyle{#1}{#2}}%
    {\@fmsl@sh\scriptscriptstyle{#1}{#2}}}
\def\@fmsl@sh#1#2#3{\m@th\ooalign{$\hfil#1\mkern#2/\hfil$\crcr$#1#3$}}
\makeatother

\title{Yang-Mills Theories on Noncommutative Space-Time}

\author{Xavier Calmet}

\address{
California Institute of Technology, Pasadena, 
California 91125, USA\\
E-mail:calmet@theory.caltech.edu}


\maketitle

\abstracts{We describe some recent progress in our understanding of
Yang-Mills theories formulated on noncommutative spaces and in
particular how to formulate the standard model on such spaces}

The idea that space-time might be noncommutative at short distances is
not new but it was taken very seriously recently because
noncommutative coordinates were found in a specific limit of string
theory. This is nevertheless not the only motivation to study
Yang-Mills theories on noncommutative spaces. In the early days of
quantum field theories, it was thought that a fundamental cutoff might
be useful to regularize the infinities appearing in these
theories. Nowadays it is understood that gauge theories describing the
strong and electroweak interactions are renormalizable and thus
infinities cancel, but it might still be useful to have a fundamental
cutoff to make sense of a quantum theory of gravity, whatever this
might be. A more pragmatic approach is that space-time could simply be
noncommutative at short distances in which case one has to understand
how the standard model can emerge as a low energy model of a
Yang-Mills theory formulated on a noncommutative space-time.

The simplest noncommutative relations one can study are $[\hat
x^\mu,\hat x^\nu] \equiv \hat x^\mu \hat x^\nu- \hat x^\nu \hat x^\mu
=i \theta^{\mu \nu}$, $\theta^{\mu\nu}\in\mathbb{C}$. Postulating such
relations implies that Lorentz covariance is explicitly broken.  These
relations also imply uncertainty relations for space-time coordinates
which are a reminiscence of the famous Heisenberg uncertainty
relations for momentum and space coordinates.  Note that $\theta^{\mu
\nu}$ is a dimensional full quantity, dim($\theta^{\mu
\nu}$)$=$mass$^{-2}$. If this mass scale is large enough, $\theta^{\mu
\nu}$ can be used as an expansion parameter like $\hbar$ in quantum
mechanics. We adopt the usual convention: a variable or function with
a hat is a noncommutative one.

There are different approaches to gauge field theory on noncommutative
spaces, see\cite{Calmet:2004dn} and references therein . If fields
are assumed to be Lie algebra valued, it turns out that only U(N)
structure groups are conceivable because the commutator $
\s[\hat\Lambda,\hat\Lambda'] = \frac{1}{2}\{\hat
\Lambda_a(x)\stackrel{\star}{,} \hat \Lambda'_b(x)\}[T^a,T^b] +
\frac{1}{2}\s[\hat \Lambda_a(x),\hat \Lambda'_b(x)]\{T^a,T^b\}$
of two Lie algebra valued noncommutative gauge parameters
$\hat\Lambda = \Lambda_a(x) T^a$ and $\hat\Lambda' = \Lambda'_a(x)
T^a$ only closes in the Lie algebra if the gauge group under
consideration is U(N) and if the gauge transformations are in the
fundamental representation of this group. But, this approach cannot be
used to describe particle physics since we know that SU(N) groups are
required to describe the weak and strong interactions. Or at least
there is no obvious way known to date to derive the standard model as a
low energy effective action coming from a U(N) group. 

Furthermore it turns out that even in the U(1) case, charges are
quantized and it thus impossible to describe quarks. This problem is
obvious if one writes the field strength tensor explicitly: $F_{\mu
\nu}=i\s[D_\mu,D_\nu]=\partial_\mu A_\nu - \partial_\nu A_\mu-i
\s[A_\mu,A_\nu]$. The commutator $\s[A_\mu,A_\nu]$ does not vanish
even for a U(1) gauge group, the choice of charges introduced in the
theory is very limited namely $\pm$1 or 0.

There is a framework that enables to address these problems. The aim
of this approach is to derive low energy effective actions for the
noncommutative theory which is too complicated to handle. The matching
of the noncommutative action to the low energy action on a commutative
space-time is done in two steps. First the noncommutative coordinates
are mapped to usual coordinates, the price to pay is the introduction
of a star product.  Secondly the noncommutative fields are mapped to
commutative fields by means of the Seiberg-Witten
maps\cite{Seiberg:1999vs}. The Seiberg-Witten maps have the remarkable
property that ordinary gauge transformations $\delta A_\mu = \pp\mu
\Lambda + i[\Lambda,A_\mu]$ and $\delta \Psi = i \Lambda\cdot \Psi$
induce noncommutative gauge transformations of the fields $\hat A$,
$\hat \Psi$: $\delta \hat A_\mu = \hat\delta \hat A_\mu$ and $\delta
\hat \Psi = \hat\delta \hat \Psi$.

The basic assumption is that the noncommutative fields are not Lie
algebra valued but are in the enveloping of the
algebra\cite{Jurco:2000ja}: $\hat\Lambda = \Lambda^0_a(x) T^a +
\Lambda^1_{ab}(x) :T^a T^b: + \Lambda^2_{abc}(x) :T^a T^b T^c: +
\ldots $, where $:\;:$ denotes some appropriate ordering of the Lie
algebra generators. One can choose, for example, a symmetrically
ordered basis of the enveloping algebra, one then has $:T^a:=T^a$ and
$:T^a T^b:= \frac{1}{2} \{T^a, T^b \}$ and so on. Taking fields in the
enveloping of the algebra allows to consider SU(N) groups. At first
sight it seems that one has introduced a infinity number of degrees of
freedom. It turns out that all fields appearing in $\hat\Lambda$ can
be expressed in terms of the classical gauge parameter. Higher order
terms in $\hat\Lambda$ are assumed to be suppressed by higher powers
of $\theta$.

Expanding to leading order in $\theta$ the star product and the
noncommutative fields, one obtains the new operators:
\begin{eqnarray*}
\label{actionqed} 
-\frac{1}{4} \theta^{\mu \nu} \bar{\psi} F_{\mu \nu} (i
\gamma^\alpha D_\alpha -m )\psi, \ -\frac{1}{2}  \theta^{\mu
\nu} \bar{\psi} \gamma^\rho F_{\rho \mu} i D_\nu \psi, \\
+\frac{1}{8} \theta^{\sigma \rho } F_{\sigma \rho }F_{\mu \nu }
F^{\mu \nu } \  \mbox{and} \ -\frac{1}{2} \theta^{\sigma \rho }
F_{\mu \sigma}F_{\nu \rho} F^{\mu \nu}.
\end{eqnarray*} 

There are a number of difficulties which have to be addressed in order
to formulate the standard model on a noncommutative space-time.  These
problems can be solved\cite{Calmet:2001na}.

The first problem is that one cannot introduce three different
noncommutative gauge potentials. The reason is that noncommutative
gauge invariance is linked to the invariance of the covariant
coordinates $\hat X^\mu = \hat x^\mu+\hat B^\mu$. The Yang-Mills
potential $A_\mu$ is related to $B^\mu$ by $B^\mu=\theta^{\mu \nu}
A_\nu$, i.e. gauge transformations are related to transformations of
the covariant coordinate. The solution is to introduce a master field:
$V_\mu=g' A_\mu+g B_\mu+g_S G_\mu$ that contains all the gauge
potential of the structure group SU(3)$\times$SU(2)$\times$U(1) and to
performed a Seiberg-Witten map for $\hat V_\mu$. Note that a
generalized gauge transformation is also introduced
$\Lambda=g'\alpha(x) Y + g \alpha_L(x)+g_s \alpha_s(x)$, with the
Seiberg-Witten map $\hat \Lambda = \Lambda +\frac{1}{4} \theta^{\mu
\nu} \{V_\nu,\partial_\mu \Lambda\}+{\mathcal O}(\theta^2)$.

The approach presented in\cite{Calmet:2001na} offers a very natural
problem to the charge quantization problem. One introduces $n$
different noncommutative hyperphotons, one for each charge entering
the model: $ \hat \delta \hat a_i^{(n)} =\partial_i \hat \lambda^{(n)}
+ i [\hat \lambda^{(n)},\hat a_i^{(n)}]$ with $\hat \delta
\hat{\Psi}^{(n)}= i e q^{(n)} \hat \lambda^{(n)} \star \hat
\Psi^{(n)}$. At first sight, it seems that this implies the existence
of $n$ photons in nature, i.e. that the theory has too many degrees of
freedom, but once again the Seiberg-Witten maps can be used to reduce
the amount of degrees of freedom. It turns out that these $n$
noncommutative hyperphotons have the same classical limit $a_i$: $\hat
a_i^{(n)}=a_i- e q^{(n)} \frac{1}{4}\theta^{kl}\{a_k,\partial_l a_i +
f_{li}\}+{\mathcal O}(\theta^2)$ i.e. there is only one classical
photon.

Another problems are the Yukawa couplings: a noncommutative field can
transform on the left-hand side or on the right-hand side and this
makes a difference. This is an obvious complication for Yukawa
couplings. For example $\bar{\hat L} \star \hat \Phi \star \hat e_R$
is not invariant under a noncommutative gauge transformation if $\hat
\Phi$ transforms only on the right-hand side or only on the left-hand
side. The solution is to assume that it transforms on both sides to
cancel the transformations of the SU(2) doublet and of the SU(2)
singlet fields, $\bar{\hat L} \star \rho_L(\hat \Phi) \star \hat e_R$
with $\rho_L(\hat \Phi)=\Phi[\phi,-\frac{1}{2}g'{\mathcal A}_\mu+g
B_\mu,g'{\mathcal A}_\mu]$ and $\hat\Phi[\Phi,A,A'] = \Phi +
\frac{1}{2}\theta^{\mu\nu} A_\nu \Big(\partial_\mu\Phi -\frac{i}{2}
(A_\mu \Phi - \Phi A'_\mu)\Big)+\frac{1}{2}\theta^{\mu\nu}
\Big(\partial_\mu\Phi -\frac{i}{2} (A_\mu \Phi - \Phi
A'_\mu)\Big)A'_\nu$. Note that it is possible to couple neutral
particles to the photon in a gauge invariant way.

Another source of model dependence originates in the choice of the
representation of the noncommutative field strength. The action for
non-Abelian noncommutative gauge bosons is $S_\mathrm{gauge} =
-\frac{1}{2}\int d^4 x \, \mbox{\bf Tr} \frac{1}{\mbox{\bf G}^2}
\widehat F_{\mu\nu} \star \widehat F^{\mu\nu}$, with the
noncommutative field strength $\widehat F_{\mu\nu}$, an appropriate
trace $\mbox{\bf Tr}$ and an operator~$\mbox{\bf G}$. This operator
must commute with all generators ($Y$, $T_L^a$, $T_S^b$) of the gauge
group so that it does not spoil the trace property of $\mbox{\bf Tr}$.
The evaluation of the trace depends on the choice of the
representation of $\widehat F^{\mu\nu}$.

 One consequence is that the triple photon vertex cannot be used to
bound space-time noncommutativity. While such an interaction can be
seen as a smoking gun of space-time noncommutativity (see
e.g.\cite{Rizzo:2002yr}), the bounds obtained are model dependent and
only constrain a combination of $\theta^{\mu\nu}$ and of an unknown
coupling constant. It is worth noting that most collider studies have
considered modifications of the gauge sector to search for space-time
noncommutativity\cite{Behr:2002wx}. While these channels and rare
decays are interesting from the discovery point of view, they cannot
be used to bound the noncommutative nature of space-time itself. The
only model independent part of the effective action is the fermionic
sector. Note that the bounds on the noncommutativity of space-time are
fairly loose if fields are taken to be in the enveloping algebra, and
are of the order of 10 TeV\cite{Calmet:2004dn}.

As a conclusion we want to emphasize the fact that models based on
Yang-Mills theories formulated on a noncommutative space-time are far
from being excluded experimentally and that a lot of progress have
been made on the theoretical side.


\begin{thebibliography}{0}


\bibitem{Calmet:2004dn}
X.~Calmet,
arXiv:hep-ph/0401097.


\bibitem{Seiberg:1999vs}
N.~Seiberg and E.~Witten,
JHEP {\bf 9909}, 032 (1999)
[arXiv:hep-th/9908142].


\bibitem{Jurco:2000ja}
B.~Jurco, S.~Schraml, P.~Schupp and J.~Wess,
Eur.\ Phys.\ J.\ C {\bf 17}, 521 (2000)
[arXiv:hep-th/0006246].


\bibitem{Calmet:2001na}
X.~Calmet, B.~Jurco, P.~Schupp, J.~Wess and M.~Wohlgenannt,
Eur.\ Phys.\ J.\ C {\bf 23}, 363 (2002)
[arXiv:hep-ph/0111115].

\bibitem{Calmet:2003jv}
X.~Calmet and M.~Wohlgenannt,
Phys.\ Rev.\ D {\bf 68}, 025016 (2003)
[arXiv:hep-ph/0305027].


\bibitem{Rizzo:2002yr}
T.~G.~Rizzo,
Int.\ J.\ Mod.\ Phys.\ A {\bf 18}, 2797 (2003)
[arXiv:hep-ph/0203240].

\bibitem{Behr:2002wx}
W.~Behr {\it et al.},
Eur.\ Phys.\ J.\ C {\bf 29}, 441 (2003)
[arXiv:hep-ph/0202121].

\end{thebibliography}
\end{document}